\documentclass[conference]{IEEEtran}
\usepackage{times}
\usepackage{helvet}
\usepackage{courier}
\usepackage[ruled,vlined]{algorithm2e}
\usepackage{graphicx}
\usepackage{url}
\usepackage{multirow}
\usepackage{booktabs}
\usepackage{amsfonts}
\usepackage{subcaption}
\usepackage{color}
\usepackage{amsmath,amssymb,amsfonts}
\usepackage{pifont}

\usepackage{ifpdf}
\ifCLASSOPTIONcompsoc

  \usepackage[nocompress]{cite}
\else
  \usepackage{cite}
\fi

\begin{document}

\title{\huge CATFL: Certificateless Authentication-based Trustworthy Federated Learning for 6G Semantic Communications}
\author{
	\IEEEauthorblockN{
		Gaolei Li \IEEEauthorrefmark{1}, 
		Yuanyuan Zhao\IEEEauthorrefmark{2}, 
		and Yi Li \IEEEauthorrefmark{3}
  }\\ 
	\IEEEauthorblockA{\IEEEauthorrefmark{1}School of Electronic Information and Electrical Engineering, Shanghai Jiao Tong University, China}
	\IEEEauthorblockA{\IEEEauthorrefmark{2}School of Information Science and Technology, Hangzhou Normal University, China \\ }
	\IEEEauthorblockA{\IEEEauthorrefmark{3} College of Information, Mechanical and Electrical Engineering, Shanghai Normal University, China\\gaolei\_li@sjtu.edu.cn, \{yyzhao04,lililiklings\}@163.com} 
} 
\maketitle
\vspace{-1em} 
\begin{abstract}
Federated learning (FL) provides an emerging approach for collaboratively training semantic encoder/decoder models of semantic communication systems, without private user data leaving the devices.
Most existing studies on trustworthy FL aim to eliminate data poisoning threats that are produced by malicious clients, but in many cases, eliminating model poisoning attacks brought by fake servers is also an important objective. In this paper, a certificateless authentication-based trustworthy federated learning (CATFL) framework is proposed, which mutually authenticates the identity of clients and server. In CATFL, each client verifies the server's signature information before accepting the delivered global model to ensure that the global model is not delivered by false servers. On the contrary, the server also verifies the server's signature information before accepting the delivered model updates to ensure that they are submitted by authorized clients. Compared to PKI-based methods, the CATFL can avoid too high certificate management overheads. Meanwhile, the anonymity of clients shields data poisoning attacks, while real-name registration may suffer from user-specific privacy leakage risks. Therefore, a pseudonym generation strategy is also presented in CATFL to achieve a trade-off between identity traceability and user anonymity, which is essential to conditionally prevent from user-specific privacy leakage. Theoretical security analysis and evaluation results validate the superiority of CATFL.

\end{abstract}

\maketitle
\quad \\

\IEEEraisesectionheading{\section{Introduction}\label{sec:introduction}}

The 6G communication goes beyond the mobile internet to embrace omnipresent Internet of Everything  applications and artificial intelligence (AI) services. This shall evolve the wireless communication networks from ``connected things" to ``connected intelligence". The 6G communications will enable the interconnections among lots of intelligent agents within a hyper-connected cyber-physical system \cite{9831429}. However, 6G communications face with many challenges generated by constrained network resources, high energy consumption, and new attack surfaces (e.g., privacy leakage, data poisoning attacks), which continues to bring significant hindrance to their world-wide realization and deployment. Semantic communication is a brand-new communication paradigm, which meets the interaction requirements of intelligent agents in the 6G era \cite{9955312,9770094,9832831}. Meanwhile, federated Learning (FL) provides a promising collaborative mode to train joint learning models for connected intelligence at the network edge \cite{DBLP:journals/corr/McMahanMRA16}. Compared to the centralized learning framework, the FL allows mobile/edge devices to collaboratively extract the distributed knowledge from the data that does not leave the devices \cite{9685654,9709603}. The characteristics of local training and joint learning of FL is helpful to protect the privacy of information senders and receivers in 6G semantic communications \cite{9685654}.
\par

\begin{figure}
  \centering
  \includegraphics[width=3.3in]{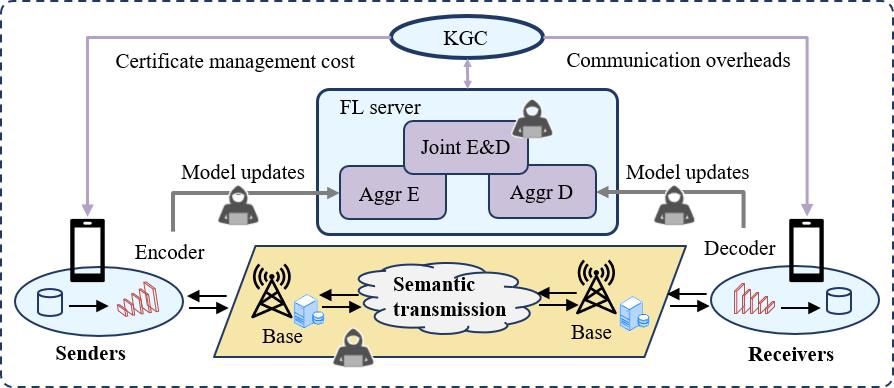}
  \caption{Security threats of federated learning in 6G semantic communications. Before participating in the joint training of semantic encoder/decoder, the client needs to obtain a certificate issued by the certification center, and only the client with the legal certificate can join the FL model aggregation. Meanwhile, before receiving the encoded semantic information sent by the sender, the receiver also needs to verify if the sender's identity is valid. 
  }
 \label{fig:Fig1}
\end{figure}

However, FL still exposes many security threats. Firstly, poisoning attacks seriously threatens its usability \cite{9772337,Yan_Poor_2021}. On one hand, since the users’ original data and the local training process is not open for the FL server, malicious FL clients may submit poisoned parameters to induce encoder/decoder errors at the model testing phase. Notably, a poisoned parameter may be diffused across all FL clients by model aggregation and distribution, thereby rendering the final global learning model to be backdoored. 
For a purpose-based terrorist to misguide other autonomous vehicles at the stop sign, it is possible to move forward by submitting maliciously-crafted inputs, leading to a serious traffic accident \cite{9501983}. Therefore, any clients and parameters they submit in the FL training process need to be authenticated and traced. On the other hand, FL is also vulnerable to model poisoning attacks because it is extremely hard to guarantee that the FL server is trustworthy and robust especially when deployed in 6G edge computing scenarios. In \cite{bagdasaryan2020how}, the attacker ambitiously substitutes the aggregated global model with a malicious model to strengthen the poisoning effect. That's why the FL server also needs to be authenticated by each client to reject false global model. 
Secondly, possible privacy leakage from FL-based semantic communication systems also should be concerned. On one hand, the semi-honest FL server can reconstruct sensitive training samples (e.g. facial images, financial records, and medical images) of the targeted FL clients through the shared gradients or parameters \cite{9666855,9546481}. On the other hand, since 6G semantic communication transmits compressed semantic information between senders and receivers, the training data can be reconstructed from the semantic representation vectors if the attacker establishes an eavesdrop mechanism on the semantic communication channels. Therefore, it is still necessary to propose an additional privacy-enhancing mechanism against such privacy leakage risks in FL-based semantic communication systems. 

Motivated by the aforementioned issues, a trustworthy FL should consider the following fundamental issues: 1) how to guarantee that users' privacy is not leaked during the model training process and 2) how to guarantee the model robustness to malicious manipulations. Therefore, resisting poisoning attacks while protecting users’ privacy has became an meaningful and urgent demand. 
Cryptography-based methods (e.g., homomorphic encryption and secret sharing) are essential to guarantee the confidentiality of shared gradients or parameters between FL clients. However, cryptography-based methods are always invalid when deployed on devices with limited computational and communication resources. For example, too high certificate management overheads reduce the practicality of PKI-based schemes in 6G semantic communication systems. 
\par
In this paper, we propose a Certificateless Authentication-based Trustworthy Federated Learning (CATFL) framework for 6G semantic communications, which can efficiently defend against poisoning attacks without leaking users' privacy. In CATFL, the FL server has two types of private key: 1) partial private key and 2) full private key, which are respectively generated by the trusted key generator center (KGC) and itself. Therefore, even an attacker who colludes with one of the CATFL server and KGC, it can not get the full secret key to impersonate, preventing the poisoning attacker from substituting the global learning model with a maliciously-modified model. Meanwhile, the CATFL is able to guarantee the trustworthiness of the local client's gradients because of providing mutual authentication for each client. We also designate a pseudonym generation strategy to hide each client's real identity. This strategy allows to trace the original real identity of each CATFL client by trusted third-parties, enabling to identify malicious mobile/edge devices. In a summary, the main contributions of our work are listed as follows:
\begin{itemize}
    \item We propose a Certificateless Authentication-based Trustworthy Federated Learning (CATFL) framework as the underlying technology of 6G semantic communications to provide higher security. Each elements in CATFL has two types of private key , which are independently generated by the KGC and itself. We prove that the CATFL can resist two types of security threats: 1) poisoning attacks (including server-side and client-side); 2) privacy leakage (including gradient leakage and semantic representation leakage).
    \item A pseudonym generation strategy is presented to achieve a trade-off between user anonymity and identity traceability in CATFL. Hence, a powerful deterrent against malicious threats to FL-based 6G semantic communication  systems is achieved. 
    \item We provide a comprehensive theoretic proof for the security and trustworthiness of proposed CATFL by comparison to existing PKI-based methods. It demonstrates that the CATFL is more applicable for practical 6G semantic communication systems.
\end{itemize}


\section{Related Work}
We comprehensively overview the security challenges of FL in 6G semantic communications, mainly caused by poisoned attacks and privacy leakage. And then, the disadvantages of existing secure authentication methods for trustworthy FL in 6G semantic communications is discussed in detail.

\subsection{Security Challenges of FL in 6G Semantic Communications}
Different from existing advanced channel encoding/decoding and modulation techniques, semantic communication attempts to extract the ``meanings” of sent information at the transmitter using artificial intelligence and then transmits these ``meanings" to the receivers. With the assistance of a shared knowledge base (KB), the receiver can accurately catch the received ``meanings" \cite{9679803}.
In 6G semantic communication systems, semantic encoder/decoder could be jointly trained using the FL framework for ensuring low loss rate of shared semantic knowledge and transmission accuracy. However, many security risks of FL have not yet been identified in details, lacking of effect countermeasures.
\par

\subsubsection{Poisoning Attacks}
Many practical poisoning attacks on FL have been proposed, which can be divided into two aspects: 1) data poisoning attacks and 2) model poisoning attacks. Usually, the data poisoning attack uses crafted data to maliciously substitute normal samples secretly. 
To generate crafted data, Zhang et al. \cite{9194010} introduced generative adversarial networks to inversely reconstruct training data from the given models, which enables a high attack efficiency. 
By model replacement and adjusting the scaling factor, Eugene et al. \cite{bagdasaryan2020how} enhanced the persistence of backdoor in FL-based systems. 
\subsubsection{Privacy Leakage Threats}
Nowadays, although the FL is designated to enhance the privacy-preserving ability, privacy leakage threats have plagued the development of FL technologies. On one hand, the untrusted parameter server may reconstruct the training data using an adversarial algorithm proposed in \cite{zhu2019deep} from the gradients submitted by each mobile/edge device. This is mainly because the updated model parameters in the training process may memory sensitive information and leak them to malicious adversaries \cite{protection2019}. 
On the other hand, the attacker can reconstruct the training data by exploiting some query-feedback information on the targeted learning model, i.e., membership inference and model inversion \cite{8835269}. 
The common feature of this kind of privacy leakage threat is to generate dummy samples that can approximate the real gradients, predictions, and weight parameters submitted by local FL clients.

\subsection{Secure Authentication for Trustworthy Federated Learning}
Secure authentication is widely deployed as an emerging technique of exchanging model parameters and certification to provide security and trustworthiness for FL-based system against privacy leakage threats and poisoning attacks. 
Authors in \cite{8765347} proposed a verifiable federated learning (VerifyNet) framework, enabling each client to verify whether the FL server works correctly. In VerfyNet, the adversary can not deceive users by forging the identity proof unless the formulated NP-hard problem is resolved. Salvia et al. \cite{Li2021SecureAF} provided an implementation case of secure authentication on the open-source FL platform named as ``Flower". 
Since there are two types of malicious behaviors: i) unauthorized manipulations to the submitted model updates; ii) sending the same model update to the FL server with multiple times, authors in \cite{Karako2022ASP} introduced data signatures to prevent from malicious behavior i). Meanwhile, this work also proposes to exploit blind signatures to sign local model updates only once to avoid malicious behavior of ii). 
However, all existing methods use the public key infrastructure (PKI)-based authentication, which is a certificate-based mechanism so that the certificate management cost and communication overheads must be affordable. Therein, many unmanned mobile/edge devices in semantic communication systems can not be equipped with such complex PKI-based authentication mechanisms. To this end, we present a novel, fundamental, efficient certificateless authentication countermeasures for deploying FL in 6G semantic communication systems. To the best of our knowledge, this is the first work to introduce certificateless authentication into the FL-based 6G semantic communication field.
\par

\section{Proposed CATFL framework}
In this section, we will introduce the proposed CATFL framework for 6G semantic communications in detail, including the preliminaries, security requirements, adversarial model and overview of proposed CATFL.

\subsection{Preliminaries}
\subsubsection{Certificateless Cryptography}
Certificateless Cryptography is first proposed by Al-Riyami et al. \cite{cryptoeprint:2003/126} to deal with the key escrow limitation of the traditional PKI-based cryptography. In Certificateless Cryptography, a trusted third party named as KGC is responsible to generate a partial private key (PSK) for the users. The user obtains the full private key by combining the PSK with a secret value, which is unknown to the KGC. Under this seeting, the KGC can not achieve the user's private keys. For FL-based 6G semantic communication systems, intelligent agents are often deployed on mobile/edge devices that has limited computation and communication resources. To reduce authentication overheads and enhance the FL trustworthiness, using Certificateless Cryptography to establish a novel 6G communication architecture has became an urgent and important demand. In this article, we will discuss its feasibility in detail (refer to Section III.D).

\subsubsection{Elliptic Curve Cryptography}
In Elliptic Curve Cryptography (ECC), there are two types of components, 1) elliptic curve, 2) pre-defined operation rules. 
Given a base point on the elliptic curve, it is very hard to calculate the discrete logarithm value of a random elliptic curve element. To save computation resouces, we will use ECC to implement the certificateless authentication mechanism in CATFL, in which the elliptic curve is formulated as the following equation:
\begin{equation}
    \mathbb{E} = \{(x,y) | y^2=x^3+ax+b, 4a^3+27b^2=0\}
\end{equation}
 
Where $\mathbb{E}$ presents an elliptic curve and $G$ is a generator with the order $r$. Given a base point $Q=kG$ and $k \in Z_r$, it is very hard and almost unpractical to find an integer $k$ in the polynomial time. In FL-based 6G semantic communication systems, we can apply the ECC to implement the certficateless authentication mechanism among the server and clients as well as the sender and the receiver.

\subsection{Security Requirements}
To bring out the motivation of our work, we summarize the security requirements of FL in 6G semantic communication systems as follows.
\begin{itemize}
     \item Message authentication: The receiver of semantic information needs to verify the semantic encodes' validity. Any tamper on the semantic encodes shall be easily detected.
     \item Non-repudiation: In CATFL, all authenticated messages could not be repudiated, that means no CATFL entities can deny a valid signature.
     \item Anonymity: Each participator needs to generate a pseudonym and its real identity has to be hidden during the communication process.
     \item Un-linkability: No CATFL entities can link multiple different messages to the same user.
     \item Resistant against attacks: The CATFL framework also needs to prevent from some typical attacks including data modification attacks and relay attacks.
     \item Conditional traceability: Only the trusted third-party that is named as KGC can know the real identity of each participator in CATFL.
\end{itemize}

\subsection{Adversarial Model}
According to listed security requirements of CATFL for 6G semantic communications, we present two types of adversaries, which are denoted as follows: 1) $\mathcal{A}_1$, and 2) $\mathcal{A}_2$. Therein, $\mathcal{A}_1$ has the ability of changing the public key of each participator using a selected constant value. However, it fails to obtain the master private key of KGC. Different from $\mathcal{A}_1$, $\mathcal{A}_2$ can get the master private key of KGC, but it fails to modify public keys of any participator.

If a valid signature $Verify (P_{pub}$, $m^*$, $AID^*$, $\Theta^*) = 1)$ is forged by an adversary $\mathcal{A}$, it can be consider the $\mathcal{A}$ ($\mathcal{A}_1$ or $\mathcal{A}_2$) has launched a successful attack. The proposed CATFL framework can be validated to be secure if the attack success rate for any attackers is negligible.

\subsection{Overview of Proposed CATFL}
The proposed CATFL framework consists of four main entities, i.e., TRA, KGC, CS and User, which connect with each other over wireless communication channels. The proposed CATFL framework has two different communication levels: 1) upper level, and 2) lower level. The upper level contains the communications between the TRA and KGC, as well as Server-to-User (S2U) communications via a secret channel, while the lower level consists of User-to-User (U2U) semantic communications via a public channel. The system entities of CATFL is shown in Fig. 2 and explained below in detail. 

\begin{itemize}
    \item Tracing Authority (TRA): The TRA in CATFL is a trusted entity with enough resources. Functions of this entity contain pseudonym generation for participants (FL server and clients) and the corresponding tracing strategy configuration (if needed). 
    \item Key Generation Center (KGC): The KGC in CATFL is an independent and trusted third party, which is responsible to distribute all the private and public keys of each participator in the 6G communication systems. 
    \item Cloud Server (CS): The CS in CATFL is responsible to receive all model updates submitted by FL clients and aggregate them (usually by averaging) to achieve an optimized global model. 
    \item Users: All message senders and receivers (also named as clients) in 6G communications, collaboratively train a uniform encoder-decoder model using the federal paradigm. Each CATFL entity trains the model with several epochs over the private local data on each mobile/edge device, and then uploads the signed model updates to CS. 
\end{itemize}

\begin{figure}
    \centering
    \includegraphics[width=3.5in]{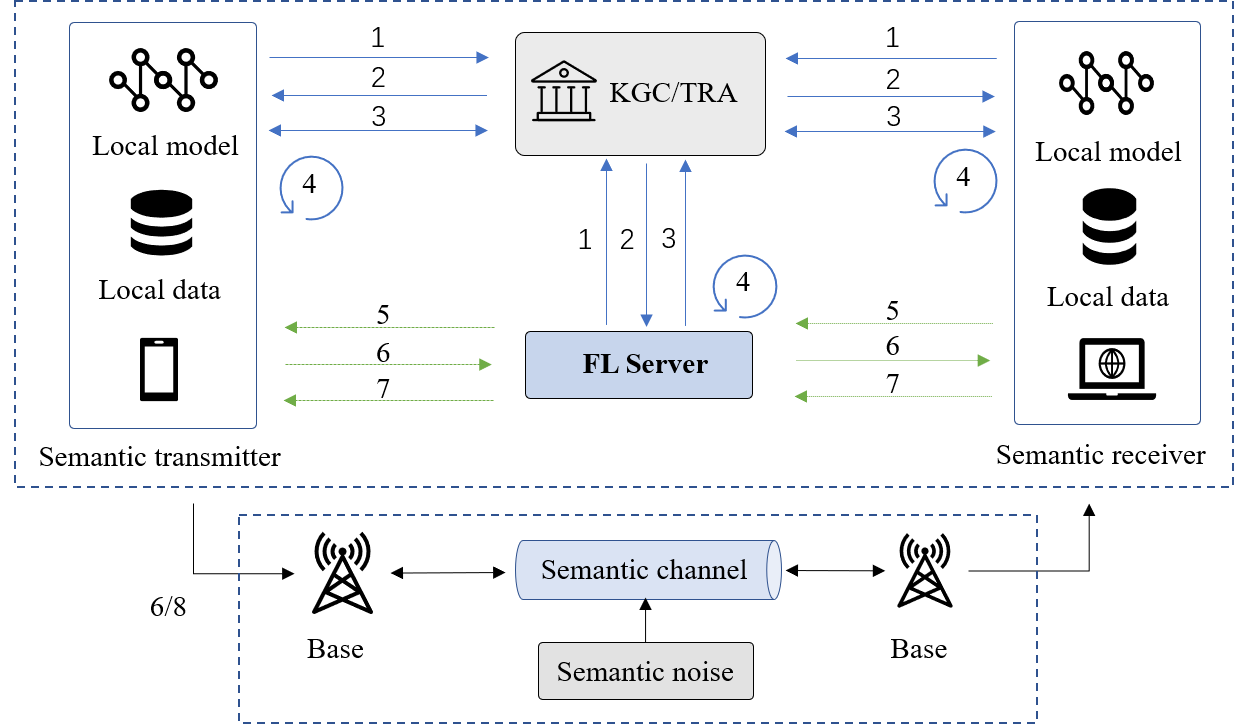}
    \caption{The workflow of proposed CATFL. 1) System setup, 2) Identity anonymization, 3) Get the PSK, 4) Extract the USK, 5) CS configuration, 6) Signature/verification, 7) Aggregation/distribution, 8) Secure semantic transmission.}
    \label{fig:overview}
\end{figure}

PKI-based methods usually involve the certificate management issues, in which a troublesome certificate revocation list should be maintained. Meanwhile, ID-based schemes may suffer from key escrow problems. We utilize certificateless cryptography to overcome these disadvantages. The concrete key generation steps in CATFL and its corresponding authentication procedures are described in detail as follows. The CATFL framework consists of eight steps in total, which are 1) System setup, 2) Identity anonymization, 3) Get PSK, 4) Extract USK, 5) CS configuration, 6) Signature/verification, 7) Aggregation/distribution, 8) Secure semantic transmission. Fig. \ref{fig:overview} illustrates the workflow of proposed CATFL framework. 

\subsubsection{System setup}
Firstly, this algorithm specifies a security parameter $\kappa$ as the input, and then it generates a cyclic additive group $(\mathbb{G}, q, P)$ and four hash functions $H_0: \mathbb{G} \times \mathbb{G} \times \{0,1\}^n \rightarrow\{0,1\}^n$, $H_1: \mathbb{G} \times \mathbb{G} \times \mathbb{G} \rightarrow Z_q^*$, $H_2: \{0,1\}^n \times \mathbb{G} \times \mathbb{G} \times \mathbb{G} \times \{0,1\}^n \rightarrow Z_q^* $ and $H_3: \{0,1\}^n \times \mathbb{G} \times \mathbb{G} \times \{0,1\}^n \rightarrow Z_q^* $, then sends $\{\mathbb{G}, q, P, H_0, H_1, H_2, H_3\}$ to TRA and KGC, respectively. After receiving those information, the TRA and KGC in CATFL will initialize system parameters according to the following three steps.
\begin{itemize}
    \item TRA in CATFL randomly selects $\alpha \in Z_q^*$, $T_{pub} = \alpha P$.
    \item The KGC in CATFL randomly selects a value $\beta \in Z_q^*$, and configures $P_{pub} = \beta P$.
    \item The information $ params = \{\mathbb{G}, q,$ $P, H_0, H_1, H_2, H_3,$ $ T_{pub},$ $P_{pub}\}$ is published.
\end{itemize}
\subsubsection{Identity anonymization}
The TRA invokes this step to initialize anonymous identities of $CS$ and each user $U_i$ in CATFL, the corresponding identity information is denoted as $RID_n$. 
\begin{itemize}
    \item $CS$ or $U_i$ in CATFL randomly selects $r_i \in Z^*_q$.
    \item $CS$ or $U_i$ in CATFL computes $AID_{i, 1} = r_i \times P$ and sends $RID_i , AID_{i, 1}$ to the TRA  where $i = 1, 2, ..., n$ through a secure wireless channel.
    \item The TRA checks the validity of $RID_i$. If not, the TRA will reject this request; Otherwise, it holds the system timestamp $T_i$ and calculates $ AID_{i, 2} = RID_i \oplus H_0(\alpha AID_{i,1}, T_{pub}, T_i)$. The pseudonym of each CATFL entity is generated as $AID_i$ = $\{AID_{i,1}, AID_{i,2}, T_i \}$.
    \item The TRA stores the computed $AID_i$ and transmits them to the other entities in CATFL.
\end{itemize}
\subsubsection{Get the PSK}
The KGC in CATFL generates and sends back the PSK to the requester secretly. After that, the requester will generate its full private key and corresponding public key for the proposed certificateless authentication mechanism. In detail, the requester $CS$ or $U_i$ sends $AID_i$ to the KGC. The KGC computes the PSK of each requester and then retrieves the $AID_i$ from the real identity list. If the $AID_i$ exists in the base, the KGC needs to execute the following operations:
\begin{itemize}
    \item Randomly produces a value $k_i \in Z^*_q$ as the system input.
    \item Calculates $U_i = k_i P$, generates $\theta_i = H_1(AID_i, U_i, $ $P_{pub})$ and then produces $\lambda_i = (k_i + \theta_i \beta)(mod \; q)$.
    \item Publishes the $\{\lambda_i, U_i\}$ to $CS$ or $U_i$ secretly.
\end{itemize}
\subsubsection{Extract the USK}
When $CS$ or $U_i$ receives $\{\lambda_i, U_i\}$, the participator will produce its key pair according to the following operations:
\begin{itemize}
    \item Calculate $\theta_i^*$ = $H_1 (AID_i , U_i , P_{pub} )$.
    \item Check whether the equation $\lambda_i P = U_i + $ $\theta_i^ * P_{pub}$ is right. If not, close the current session; Otherwise, go to the next step.
    \item Computes $X_i = \mu_iP$ and configures the public key as $PK_i = \{X_i, U_i\}$. Subsequently, these public keys will be shared with other CATFL entities.
\end{itemize}

In CATFL, a batch of $AID_i$ and PSK $\{\lambda_i, U_i\}$ will be pre-loaded into the mobile/edge devices and store safely. Each user can utilize a unique $AID_i$ and a PSK $\{\lambda_i, U_i\}$ to validate the identities of other entities. If the user uses up every $AID_i$, it can build a new connection with the TRA again and replenish a stock of $AID_i$ and $\{\lambda_i, U_i\}$ using a secure communication channel.

\subsubsection{CS configuration}
The CS in CATFL will serve as a parameter server to provide model aggregation services. First, the server needs to initialize the weights of the global model and required model parameters (e.g., the total number of FL rounds, the total number of FL clients, and the participation rate of FL clients in each training round). It then activates all selected FL clients and broadcasts the initialized global model for local training.

\subsubsection{Signature/verification}
The signature process is invoked by any $CS$ or $U_i$ to compute message/signature pairs, $CS$ or $U_i$ will execute the following operations:
\begin{itemize}
    \item Randomly chooses $a_i \in Z^*_q$, and then generate the message signature by computing $A_i = a_iP$, $h_{1, i} $ $= H_2(m_i, AID_i, PK_i, A_i, P_{pub}, t_i)$, $h_{2, i} = H_3(m_i, AID_i, $ $PK_i, A_i, P_{pub}, h_{1, i})$. Therein, $t_i$ denotes the timestamp of the whole system.
    \item Calculates $\eta_i = a_i - h_{1,i}\mu_i - h_{2,i}\lambda_i (mod q)$.
    \item Configures the message signature as $\Theta_i = \{\eta_i , A_i \}$ and broadcasts $(m_i, AID_i, \theta_i,PK_i, \Theta_i, t_i)$ to other relational $CS$ or $U_i$.
\end{itemize}

The identity verification process is also invoked by any $CS$ or $U_i$, which aims to verify the validity of each CATFL entity. If yes, the receiver in CATFL can accept the semantic information and perform further actions. The identity verification process is shown below:
\begin{itemize}
    \item Verifies the parameters $T_i$ and $t_i$. If both of them are not fresh, the received semantic information will be discarded.
    \item Checks if $\Theta_i$ is equal to $H_1(AID_i, U_i, P_{pub})$. If not, discards this traffic status; Otherwise, continues to execute further operations.
    \item Computes the hash values one by one $h_{1, i}^*$ = $H_2 (m_i,$ $AID_i, PK_i, $ $ A_i, P_{pub}, t_i)$, and $h_{2, i}^*$ = $H_3(m_i, AID_i,$ $PK_i, A_i, P_{pub}, h^*_{1,i})$ and $A_i^*$ = $\eta_i P + $ $h_{1, i}^{*} X_i + h^{*}_{2, i} $ $ U_i + (h^{*}_{2, i} $ $\theta_i)P_{pub}$ respectively.
    \item Validates if the value of $A_i$ is equal to $A^*_i$. If not, this semantic information will be discarded; Otherwise, the traffic status can be acceptable.
\end{itemize}

Since the CATFL is derived from \cite{9762546}, the Proof of Correction can refer to that article.

\subsubsection{Aggregation/distribution}
The CATFL server first aggregates the model updates sent by each FL client and then sends back the aggregated global model to the FL clients for the next training round.

\section{Security Analysis and Evaluation}
The CATFL can be proved to meet the aforementioned security requirements. Specifically, to prove the security of proposed CATFL, we design two types of games: 1) playing between a challenger $\mathcal{C}$ and the adversaries $\mathcal{A}_1$, 2) playing between a challenger $\mathcal{C}$ and the adversaries $\mathcal{A}_2$.

The detailed security analysis is shown as follows:
\begin{enumerate}
    \item Message Authentication/Integrity: According to the features of ECC, the CATFL is proven secure under two types of adversarial models so that the FL model updates and transmitted semantics can not be forged. 
    \item Anonymity: The anonymous identity 
    is used to hide the real identity of CATFL entities because the attacker can not compute $\alpha AID_{i, 1}$. 
    \item Non-Repudiation: 
    The TRA could trace the real identity $RID_i$ of each CATFL entity through the pseudonym identity $AID_i$. Therefore, in the proposed CATFL, no entity can deny the validity of the received signature.
    \item Conditional Traceability: 
    The real identity $RID_i$ can be reconstructed only by the TRA via its master secret key $\alpha$. If poisoned model updates are submitted by a malicious client, the TRA can trace this client. 
    \item Server impersonation attack: To persistent the backdoor in FL encoder/decoder model of 6G semantic communications, the attacker should generate a message/signature pair $\{m_i, AID_i, \theta_i, PK_i, \Theta_i, t_i\}$, satisfying the equations: 
    \begin{equation}
        \theta_i = H_1(AID_i, U_i, P_{pub})
    \end{equation}
    \begin{equation}
        A_i = \eta_i P + h_{1,i}X_i + h_{2,i}(U_i + \theta_i P_{pub})
    \end{equation}
    However, $\mathcal{PPT}$ attackers cannot generate the valid message/signature pair because none of them can solve the puzzle in ECC. To this end, the CATFL can successfully resist the server impersonation attack.
    \item Un-Linkability: In the identity anonymization phase, the TRA randomly picks $r_i$ to generate a pseudonym for the requester. In addition, the requester randomly selects $a_i$ to generate the message signature. It is not possible for $\mathcal{PPT}$ attackers to connect any two pseudonym identities or two signatures to a specific user.
    \item Modification attack: If a $\mathcal{PPT}$ attackers modified the message/signature pair $\{m_i, AID_i, \theta_i, PK_i, \Theta_i\}$, the modification could be discovered by checking the equations $\theta_i = H_1(AID_i, U_i, P_{pub})$ and $A_i = \eta_i P + h_{1, i} X_i $ $ + h_{2, i}(U_i + \theta_i P_{pub})$. Therefore, our CATFL framework can resist the modification attack.
\end{enumerate}

\begin{figure}
    \centering
    \includegraphics[width=3.4in]{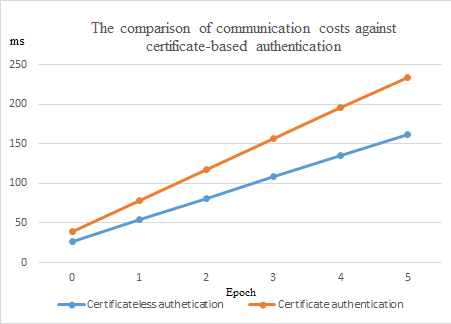}
    \caption{The comparison of communication costs against certificate-based authentication.}
    \label{fig:comparison}
\end{figure}

To bring out the efficiency of proposed CATFL, we compare the communication latency with existing methods. Firstly, since the CATFL contains two stages: 1) training stage and 2) testing stage, the communication latency for both two stages are computed, respectively. The signature latency for each training round is denoted as $T_{sign}$, and the corresponding verification latency is presented as $T_{veri}$. Thus, for $N$ training rounds, the communication cost of CATFL is $N * (T_{sign} + T_{veri})$. For $M$ messages exchanged between the sender and the receiver, the communication cost is $M * (T_{sign} + T_{veri})$. As an instance, Fig. \ref{fig:comparison} shows the comparison of communication costs against certificate-based authentication. Another impact factor for the communication cost is the number of CATFL entities $K = 2*P+1$, where $P$ is the number of sender/receiver pairs. Besides, since the FL-based system should obey the synchronous principle, it exists a waiting latency that obeys the Poisson distribution: $\Delta T_{ca} \sim \pi(\lambda)$. 

\section{Conclusion}
In this paper, we propose a Certificateless Authentication-based
Trustworthy Federated Learning (CATFL) for 6G Semantic Communications scheme. With the certificateless authentication technique, the CATFL entities have two types of private key (i.e., partial and full), which are generated independently by the trusted authority and itself. Therefore, even an attacker who colludes with the KGC cannot obtain the participant’s full secret key. On the basis of this, the proposed CATFL can prevent the semi-honest servers from inferring the users' private data. The security analysis and evaluation shows the proposed CATFL has higher security and lower communication cost against poisoning attacks and privacy leakage threats. In the future, we also can introduce more emerging techniques (e.g., blockchain) to construct a secure and efficient 6G semantic communication systems.

\ifCLASSOPTIONcompsoc
  \section*{Acknowledgments}
\else
  \section*{Acknowledgment}
\fi
This research work is funded by Shanghai Sailing Program under Grant No. 21YF1421700, Defence Industrial Technology Development Program Grant No. JCKY2020604B004, and National Nature Science Foundation of China under Grant No. 62202303 and U20B2048.

\bibliographystyle{IEEEtran}
\bibliography{ref.bib}


\end{document}